\documentclass[%
 preprint,
 amsmath,amssymb,
 aps,
 prb,
]{revtex4-2}

\usepackage{graphicx}
\usepackage{dcolumn}
\usepackage{bm}
\usepackage{subfig}
\usepackage[colorlinks=true,urlcolor=blue,citecolor=blue,linkcolor=blue,breaklinks=true]{hyperref}
\usepackage{caption}
\newcommand{\Tr}{\operatorname{Tr}}
\newcommand{\sech}{\operatorname{sech}}

\begin{document}


\title{Field theoretical approach to spin models}

\author{Feng Liu}
\affiliation{School of Physics, Peking University, Beijing 100871, China}
\affiliation{Collaborative Innovation Center of Quantum Matter, Beijing,
China}

\author{Zhenhao Fan}
\affiliation{School of Physics, Peking University, Beijing 100871, China}
\affiliation{Collaborative Innovation Center of Quantum Matter, Beijing,
China}

\author{Zhipeng Sun}
\affiliation{School of Physics, Peking University, Beijing 100871, China}
\affiliation{Collaborative Innovation Center of Quantum Matter, Beijing,
China}

\author{Xuzong Chen}
\affiliation{School of Electronics Engineering and Computer Science,
 Peking University, Beijing 100871, People’s Republic of China}

\author{Dingping Li}
\email{lidp@pku.edu.cn}
\affiliation{School of Physics, Peking University, Beijing 100871, China}
\affiliation{Collaborative Innovation Center of Quantum Matter, Beijing,
China}

\date{\today}

\begin{abstract}
  We developed a systematic non-perturbative method base on Dyson-Schwinger theory and
  the $\Phi$-derivable theory for Ising model at broken phase.
  Based on these methods, we obtain critical
  temperature and spin spin correlation beyond mean field theory. The spectrum of Green
  function obtained from our methods become gapless at critical point, so the
  susceptibility become divergent at $T_c$. The critical temperature of Ising model
  obtained from this method is fairly good in comparison with other non-cluster methods.
  It is straightforward to extend this method to more complicate spin models
  for example with continue symmetry.
\end{abstract}

\maketitle


\section{\label{sec:intro}Introduction}
The Ising model is the simplest spin model, and it has been studied for many years.
Rigorous solutions have been given by Ising for the one-dimensional
case\cite{ising1925beitrag} and by Onsager in the case of the two-dimensional
square lattice\cite{onsager1944crystal}, which provide a benchmark for other
approximate method. There are several techniques for solving such kind statistical models, such
as Monte Carlo simulations, mean-field-type methods, cluster mean-field methods,
and renormalization-group methods (RG). Although RG methods have a good description
of system in the vicinity of the critical point, it neither predicts the behavior of
the system in the area far away from critical point nor give the phase transition
temperature. Many attempts to calculate quantities in the statistical systems
beyond the mean field theory have been made in recent years \cite{kuzemsky2009statistical}.

The mean-field (MF) approach, base on one-site approximation, began from
Pierre Weiss \cite{weiss1907magnetism}, which gives the well-know solution for the
critical temperature of the transition from symmetry phase to broken phase $T_c/J=z$,
where $z$ is the number of nearest neighbors and $J$ is the coupling strength.
Wysin and Kaplan \cite{wysin2000correlated} made a significant improvement to the
MF in a simple way. Their ``self-consistent correlated molecular-field theory" (SCCF)
take into account the impact of the spin state of the central spin on the effective
field of neighboring spins. They obtained more accurate critical temperature
compared with some other methods, such as MF or Bethe-Peierls-Weiss (BPW) approximation (often called Bethe
approximation in short)\cite{bethe1935statistical,peierls1936ising,weiss1948application}.
Zhuravlev \cite{zhuravlev2005molecular} introduce ``screened magnetic field"
approximation which further improves the result of the SCCF method and allows
one to obtain critical temperature with better accuracy. Beyond on-site approximation,
BPW approach can be considered as the simplest case of
cluster approach. Base on cluster idea, many new approximations have been proposed,
such as ``correlated cluster mean-field" (CCMF) theory introduced by
Yamamoto \cite{yamamoto2009correlated}, ``effective correlated
mean-field approach" (ECMF) developed by Viana \cite{viana2014effective}.
For a large enough cluster, this approach can give a good estimate for critical
temperature.

In this work, we use two kind field theoretical methods to treat Ising model, which is
based on Schwinger-Dyson equations (1PI) approach and two-particle irreducible (2PI)
$\Phi$-derivable theory \cite{luttinger1960ground,baym1961conservation,
cornwall1974effective}, respectively. Within the approximations, both methods
do not necessarily guarantee the identity which respects the fluctuation-dissipation theorem,
thus the susceptibility of Ising model obtained from above methods do not diverge
at critical temperature. Fortunately,
for 1PI approach, a general method to preserve the identity in an approximation
scheme was developed long time ago \cite{kovner1989covariant} in the context of field
theory as covariant Gaussian approximation(CGA) to solve unrelated problems in quantum
field theory and superfluidity. For 2PI method, Van Hees and Knoll developed an
improved $\Phi$-derivable theory which preserve the identity by approximating the
1PI functional with the 2PI functional\cite{van2002renormalization}.  After modified
procedure mentioned above, the susceptibility of Ising model diverges at critical
temperature for both case. With a relatively low cost, the critical temperature $T_c$
obtained from them is quite accurately comparing with other non-cluster method.
More importantly, since our methods base on Hubbard-Stratonovich transformation,
it is straightforward to extend these methods to more complicate models, like XY model,
Heisenberg model, with preserving the identity for the fluctuation-dissipation theorem and Ward-Takahashi
identity (WTI) for models with continue symmetries, which are crucial for the description of such systems.

The paper is organized as following. In Sec.\ref{sec:1pi} and Sec.\ref{sec:2pi},
we derive the
equations for Ising model base on 1PI approach and the $\Phi$-derivable theory
respectively. Numerical results including the critical temperature, susceptibility
and Green's function are given in Sec.\ref{sec:res}. Finally, we give a summary
in Sec.\ref{sec:con}.

\section{\label{sec:1pi}1PI formalism}
The Hamiltonian of the Ising model in a two-dimension square lattice can
be expressed as
\begin{equation}
  H =-\frac{1}{2}\sum_{i,j}J_{i,j}\sigma_{i}\sigma_{j}-\sum_{i}\sigma_{i}h_{i}
\end{equation}
where $J_{i,j}$ is the coupling strength between  $i$ and $j$, which equal to $J$ for any two
nearest neighboring  sites, otherwise equal to zero. The spin $\sigma_i$ takes either $+1$ or $-1$.

After Hubbard-Stratonovich transformation, the grand-canonical partition function
of this system can be written as path integral over continue variable parameter
$\phi$ \cite{amit2005field}:

\begin{widetext}
\begin{equation}
  Z[h] =\sum_{\{\sigma_i\}} \exp\left[-\beta H\right]
       =\int D[\phi] \exp\left(-\frac{\beta}{2}\sum_{i,j}J^{-1}_{i,j}
       \left(\phi_{i}\phi_{j}-h_{i}\phi_{j}-\phi_{i}h_{j}+h_{i}h_{j}\right)
       +\sum_{n} \ln [\cosh[\beta\phi_{n}] ]\right)
\end{equation}
\end{widetext}

Based on the above formula, we can get the relationship between $\sigma$ and $\phi$
for zero-external field case i.e. $h_{i} = 0$ for each $i$.

\begin{equation} \label{eq:defsigma}
  \left<\sigma_{m}\right>=\sum_{i} J^{-1}_{m,i}\left<\phi_i\right>
\end{equation}

\begin{equation}
  \left<\sigma_{m}\sigma_{n}\right>_{c}= \sum_{i,j}J^{-1}_{m,i}J^{-1}_{n,j} \left<\phi_{i}\phi_{j}\right>_{c}
- \beta^{-1} J^{-1}_{m,n}
\end{equation}
here we have used the property $J_{i,j} = J_{j,i}$, thus $J^{-1}_{i,j} = J^{-1}_{j,i}$.

For zero-external field case, we add a new auxiliary source $H_i$ to generate
Green function. This auxiliary source has to be set to zero at the end of the
calculation. The partition function can be rewritten as:

\begin{equation}
  Z\left[H\right]=\int D\left[\phi\right] \exp\left(-\beta\frac{1}{2}\sum_{i,j}
    J^{-1}_{ {i} {j}}\phi_{ {i}}\phi_{ {j}}
    +\sum_{ {n}}
    \ln\left[\cosh\left[\beta\phi_{ {n}}\right]\right]-\sum_{i}H_{ {i}}\phi_{ {i}}\right)
\end{equation}
The generating functional $W$ for connected diagrams reads $W\left[H\right]=-\ln Z\left[H\right]$,
From this we can define the mean field and the connected Green's function:

\begin{equation}
    \varphi_{ {i}} =\frac{\delta W\left[H\right]}{\delta H_{ {i}}}=\left<\phi_{ {i}}\right>
\end{equation}

\begin{equation}
  G_{ij} =-\frac{\delta^2W[H]}{\delta H_i \delta H_j}=\left<\phi_i \phi_j \right>-\left<\phi_i \right>\left< \phi_j \right>
\end{equation}

By a functional Legendre transformation on $\varphi$ one obtains the effective action:

\begin{equation}
  \Gamma\left[\varphi\right]=W\left[H\right]-\sum_{i}H_{{i}}\varphi_{{i}}
\end{equation}

The first equation in the series of the DS equations, i.e., the off-shell($H \neq 0$) ``shift" equation is
\begin{equation} \label{eq:shift}
  0 = H_{{m}}+\beta \sum_{i}J^{-1}_{{i}{m}}\varphi_{{i}}-\beta\left<\tanh\left[\beta\phi_{{m}}\right]\right>
\end{equation}
Higher-order DS equations in the cumulant form are obtained
by differentiating the equation above. The second DS equation is

\begin{equation} \label{eq:gap}
  \Gamma_{ij}=-\frac{\delta H_{i}}{\delta \varphi_{j}}=\beta J^{-1}_{{ij}}
-\beta\frac{\delta\left<\tanh\left[\beta\phi_{{i}}\right]\right>}{\delta \varphi_{j}}
\end{equation}

$\Gamma_{ij}$ is the inverse of $G_{ij}$ since
\begin{equation}
 \sum_{n} G_{in}\Gamma_{n j}=\sum_{n}\frac{\delta^2W[H]}{\delta H_i\delta H_n}\frac{\delta H_n}
  {\delta\varphi_j}=\sum_{n}\frac{\delta H_n}{\delta\varphi_j}\frac{\delta \varphi_i}{\delta H_n}
=\delta_{ij}
\end{equation}
Consider leading correction to mean field theory, the $\left<\tanh\left[\beta\phi_{{i}}\right]\right>$ can
be expanded as
\begin{equation} \label{eq:tanh}
  \left<\tanh[\beta\phi_i]\right>
  =\tanh[\beta\varphi_i]-\beta^{2}\sech\left[\beta\varphi_i
  \right]^{2}\tanh\left[\beta\varphi_i\right]G_{ii}
\end{equation}
Substitute Eq(\ref{eq:tanh}) into Eq (\ref{eq:shift}) and (\ref{eq:gap}), and neglect the
derivative of $G_{ii}$ with respect to $\varphi_{j}$ according to leading order approximation.
Now we could set $H=0$, for homogeneous system we have $\varphi_i=\varphi$ for
any site $i$. Thus, we could express above equations in momentum space using Fourier
transformation $G_{ij}=\sum_{\alpha=x,y}\int^{\pi}_{-\pi} \frac{d^2k}{(2\pi)^2} \exp(-ik_\alpha(i_\alpha-j_\alpha))G(k)$:

\begin{equation} \label{eq:1pishift}
  0=\frac{\beta \varphi}{4J}-\beta
    \tanh\left[\beta \varphi\right]+\beta^{3}G_{ii}\sech\left[\beta \varphi\right]^{2}
    \tanh\left[\beta \varphi\right]
\end{equation}

\begin{equation} \label{eq:1pigap}
  \begin{aligned}
  \Gamma(k) =& \beta J^{-1}(k)-\beta^{2}\sech\left[\beta \varphi\right]^{2} \\
   &+\beta^{4}G_{ii} \left( \sech\left[\beta \varphi\right]^{4}-2
   \sech\left[\beta \varphi\right]^{2}\tanh\left[\beta \varphi\right]^{2}\right)
  \end{aligned}
\end{equation}
Where $J^{-1}(k)=\left[2J\left(\cos(k_{x})+\cos(k_{y})\right) \right]^{-1}$, $G_{ii}=\int^{\pi}_{-\pi} \frac{d^2k}{(2\pi)^2}\ G(k)
=\int^{\pi}_{-\pi} \frac{d^2k}{(2\pi)^2} \Gamma^{-1}(k)$.
Notice $G_{ii}$ is not a function of $i$ due to translation invariance of system.
Then we can get $\varphi$ and $G$ at fixed $\beta$ and $J$ with Eq(\ref{eq:1pishift}) and Eq(\ref{eq:1pigap}).

The susceptibility obtained from above calculation do not diverge at phase
transition temperature since the truncation applied to the formula (\ref{eq:tanh})
and (\ref{eq:1pigap})
will break the fluctuation-dissipation theorem. It's not surprising since such
method do not respect also WTI for systems with continue symmetry and there isn't Goldstone
modes for broken phase \cite{wang2017covariant}.

A general method to preserve both identities in an approximation scheme was developed
long time ago \cite{kovner1989covariant, rosenstein1989covariant}, in the context of field theory
as the covariant Gaussian approximation(CGA). In this improved method, the full covariant correlator
is defined by functional derivative:

\begin{equation}
  \begin{aligned}
    (G_{\text{full}})^{-1}_{ij}&=-\frac{\delta H_{i} }{\delta  \varphi_{j}}\\
    &=\Gamma_{ij}+\beta^{3}\Lambda_{iij}\sech\left[\beta \varphi_i \right]^{2}\tanh\left[\beta \varphi_i \right]
  \end{aligned}
\end{equation}
where

\begin{equation}
  \Lambda_{iij}=\frac{\delta G_{ii}}{\delta \varphi_j}
\end{equation}
which can be obtained by taking the derivative of
\begin{equation}
  \delta_{ij}=\sum_{n}G_{in}\Gamma_{n j}
\end{equation}

we get

\begin{equation} \label{eq:bse}
  \Lambda_{iim}=-\sum_{k,j} G_{ik}\frac{\delta \Gamma_{kj}}{\delta \varphi_{m}}G_{ji}
\end{equation}
here
\begin{equation}
  \begin{aligned}
    \frac{\delta \Gamma(i,j)}{\delta \varphi_m}=&
      -\beta^{2}\frac{\delta  (\text{sech}\left[\beta\varphi_{ {i}}\right]^{2})}{\delta\varphi_{m}}\delta_{ {i}{j}}
      +\beta^{3}G_{ii}\frac{\delta\left(\beta  \text{sech}\left[\beta\varphi_{ {i}}\right]^{4}-2\beta
      \text{sech}\left[\beta\varphi_{ {i}}\right]^{2}\tanh\left[\beta\varphi_{ {i}}\right]^{2}\right)}{\delta \varphi_m}\delta_{ {i}{j}}\\
      &+\beta^{3}\Lambda_{iim}\left(\beta  \text{sech}\left[\beta\varphi_{ {i}}\right]^{4}-2\beta
       \text{sech}\left[\beta\varphi_{ {i}}\right]^{2}\tanh\left[\beta\varphi_{ {i}}\right]^{2}\right)\delta_{ {i}{j}}
  \end{aligned}
\end{equation}

These equations are actually the Bethe-Salpeter equation.
After Fourier transformation, We can get full covariant correlator:

\begin{equation} \label{eq:fullcorr}
  G^{-1}_{\text{full}}(k)=\Gamma(k)
 +\beta^{3}\Lambda(k)\sech\left[\beta\varphi\right]^{2}\tanh\left[\beta\varphi\right]
\end{equation}
where $\Lambda(k)$ is the Fourier transform of $\Lambda_{iim}$
\begin{equation}
  \Lambda_{iim}=\sum_{\alpha=x,y}\int^{\pi}_{-\pi}\frac{d^2k}{(2\pi)^2}\exp(-ik_\alpha(i_\alpha-m_\alpha))\Lambda(k)
\end{equation}
which can be solved by the Bethe-Salpeter equation(\ref{eq:bse}).

\begin{widetext}
  \begin{equation} \label{eq:Lambda}
    \Lambda(k)=-\frac{I(k)[\left(2\beta^{3} \sech[\beta \varphi]^2 \tanh[\beta \varphi]\right)
    +\beta^{3}G_{ii}\left(-8\beta^{2} \sech[\beta \varphi]^4 \tanh[\beta \varphi]+4\beta^2  \sech[\beta \varphi]^2 \tanh[\beta \varphi]^3\right))]}{1+I(k) \left(\beta^4 \sech\left[\beta\varphi\right]^{4}-2\beta^4 \sech\left[\beta\varphi\right]^{2}\tanh\left[\beta\varphi\right]^{2}\right)}
  \end{equation}
\end{widetext}
where $I(k)$ is defined as
\begin{equation}
  I(k)=\frac{1}{(2\pi)^{2}}\int^{\pi}_{-\pi} d^2p G(k+p)G(p)
\end{equation}

Substituting Eq(\ref{eq:Lambda}) back to Eq(\ref{eq:fullcorr}), we can get full covariant
Green's function. And the susceptibility obtained by this method will diverge at phase transition point.

\section{\label{sec:2pi}2PI formalism}
The $\Phi$-derivable approximation possesses several intriguing features.
Approximations of this kind are the so-called conserving approximations \cite{baym1961conservation,van2002renormalization}, which means it is consistent with the conservation laws that follow from the Noether's theorem (current conservation, total momentum, total energy, etc). The usual thermodynamic relations between pressure, energy density
and entropy hold exactly within this approximation.

In addition to the usually introduced one-point auxiliary
external source a two-point auxiliary external source is also
included in 2PI method. The corresponding grand-canonical partition function  is defined within
the path integral formalism as

\begin{equation}
    Z\left[H,B\right]=\int D\left[\phi\right] \exp\left(-S[\phi]-\sum_{i}H_{{i}}
    \phi_{{i}}-\frac{1}{2}\sum_{i,j}B_{ij}\phi_i\phi_j\right)
\end{equation}

where

\begin{equation}
  S[\phi]=\beta\frac{1}{2}\sum_{i,j}J^{-1}_{{i}{j}}\phi_{{i}}\phi_{{j}}
  -\sum_{{n}}\ln\left[\cosh\left[\beta\phi_{{n}}\right]\right]
\end{equation}

The generating functional of connected Green function is defined as
\begin{equation}
  W[H,B]=-\ln Z[H,B]
\end{equation}

The 2PI functional $\Gamma[\varphi, G]$ is
defined by the double Legendre transformation and can be
written in the form

\begin{equation}
  \Gamma[\varphi,G]=S[\varphi]+\frac{1}{2}
\Tr[D^{-1}\left(G-D\right)]
+\frac{1}{2}\Tr\ln(G^{-1}) + \Phi [\varphi, G]
\end{equation}
where $\left(D^{-1}\right)_{ij} = \frac{\delta^2S[\varphi]}
{\delta \varphi_i \delta \varphi_j}$ and $\varphi_i=\frac{\delta W[H,B]}{\delta H_i}
=\left<\phi_i\right>$, and $G_{ij}=\left<\phi_i \phi_j \right>-\left<\phi_i \right>\left< \phi_j \right>
$. $\Phi[\varphi, G]$ can be calculated approximately with well-known standard techniques
\cite{van2002renormalization}. We generalize   the  work \cite{van2002renormalization} by Van Hees and
J. Knoll to arbitrary interaction form, and we find the lowest order approximation of $\Phi[\varphi,G]$ can be
demonstrated to be equal to:
\begin{equation}
\Phi[\varphi,G]=\frac{1}{8}\sum_{i}S^{(4)}[\varphi_i]G_{ii}G_{ii}
\end{equation}
here $S^{(4)}[\varphi]$ stands for the fourth derivative of $S^{(4)}[\varphi]$.
The above expression will allow us obtaining the  results with O(N) Linear-Sigma model in
Ref. \cite{van2002renormalization}, but for our case:
\begin{equation}
  \Phi[\varphi, G] = \frac{ \beta^4}{4} \sum_{i}(G_{ii})^2(\sech[\beta \varphi_i]^
  {4}-2
  \sech[\beta\varphi_i ]^{2} \tanh[\beta\varphi_i]^{2} )
\end{equation}

Then the equations  are now given by the fact
that we wish to study the theory with vanishing auxiliary sources $H$ and $B$.

\begin{equation} \label{eq:2pidgdp}
  \frac{\delta \Gamma[\varphi,G]}{\delta \varphi_i}=-H_i-\frac{1}{2}\sum_{m}B_{im}
  \varphi_m-\frac{1}{2}\sum_{m}B_{mi}\varphi_m\overset{!}{=}0
\end{equation}

\begin{equation} \label{eq:2pidgdg}
  \frac{\delta \Gamma[\varphi,G]}{\delta G_{ij}}=-\frac{1}{2}B_{ij}\overset{!}{=}0
\end{equation}

Then from Eq(\ref{eq:2pidgdp}) and Eq(\ref{eq:2pidgdg}) we get the ``shift" equation
and gap equation:
\begin{equation}
  \begin{aligned}
    0=&\beta \sum_{j}J_{ij}^{-1}\varphi_j-\beta \tanh[\beta \varphi_i]+G_{ii}\left(\beta^3 \text{sech}[\beta \varphi_i]^{2}\tanh[\beta \varphi_i]\right)\\
   &-\beta^5G_{ii}G_{ii}(2\text{sech}[\beta \varphi_i]^{4}\tanh[\beta \varphi_i]
   -\text{sech}[\beta \varphi_i]^{2}\tanh[\beta \varphi_i]^3 )
  \end{aligned}
\end{equation}

\begin{equation}
  G_{ij}^{-1}=\beta J_{ij}^{-1}-
\delta_{ij}(\beta^2 \text{sech}[\beta \varphi_i]^2)
+\delta_{ij}\beta^4 G_{ii}( \text{sech}[\beta \varphi_i]^4-2\text{sech}[\beta\varphi_i ]^{2} \tanh[\beta\varphi_i]^{2} )
\end{equation}

and in Fourier space the equations reads:

\begin{equation} \label{eq:2pishift}
  \begin{aligned}
    0=&\frac{\beta \varphi}{4J}-\beta
    \tanh\left[\beta \varphi\right]+\beta^{3}G_{ii}\sech\left[\beta \varphi\right]^{2}
    \tanh\left[\beta \varphi\right]\\
    &-\beta^5G_{ii}G_{ii}(2\sech[\beta \varphi]^{4}
    \tanh[\beta \varphi]-\sech[\beta \varphi]^{2}\tanh[\beta \varphi]^3 )
  \end{aligned}
\end{equation}

\begin{equation} \label{eq:2pigap}
  \begin{aligned}
  G^{-1}(k) =& \beta J^{-1}(k)-\beta^{2}\sech\left[\beta \varphi\right]^{2} \\
   &+\beta^{4}G_{ii}\left( \sech\left[\beta \varphi\right]^{4}-2
   \sech\left[\beta \varphi\right]^{2}\tanh\left[\beta \varphi\right]^{2}\right)
  \end{aligned}
\end{equation}
we can get $\varphi$ and $G$ from Eq(\ref{eq:2pishift}) and Eq(\ref{eq:2pigap}) at fixed $J$
and $\beta$. In general the solution of Eq(\ref{eq:2pishift}) and Eq(\ref{eq:2pigap}) do not
respect symmetry of system for truncated $\Phi[\varphi, G]$. In order to cure this
problem we supplement the 2PI approximation scheme by an additional effective action
defined with respect to the self-consistent solution as \cite{van2002renormalization}
\begin{equation}
  \Gamma[\varphi] = \Gamma[\varphi, \tilde{G}[\varphi]]
\end{equation}
where $\tilde{G}[\varphi]$ is defined by
\begin{equation*}
  \left. \frac{\delta \Gamma[\varphi, G]}{\delta G} \right|_{G = \tilde{G}[\varphi]} = 0
\end{equation*}
We can define external Green's function by the usual definition as double
derivatives of $\Gamma[\varphi]$ as
\begin{equation} \label{eq:mode2pi}
  (G_{\text{ext}})^{-1}_{ij}=\frac{\delta^2\Gamma[\varphi]}{\delta\varphi_i\delta\varphi_j}
  =G_{ij}^{-1}+\frac{\delta\Phi[\varphi,G]}{\delta\varphi_i\delta\varphi_j}
  +\sum_{m,n}\frac{\delta^2\Gamma[\varphi,G]}{\delta\varphi_i\delta G_{mn}}\Lambda_{mnj}
\end{equation}
where $\Lambda_{mnj}=\frac{\delta G_{mn}}{\delta \varphi_j}$. And
\begin{equation}
  \begin{aligned}
    \frac{\delta ^2 \Gamma[\varphi,G]}{\delta \varphi_i \delta G_{mn}}
  =&\delta_{in}\delta_{mn}\left[(\beta^3 \sech[\beta \varphi_i]^2\tanh[\beta\varphi_i])
  \right. \\
  &-G_{mm}\left(4\beta^5\sech[\beta \varphi_i]^{4}\tanh[\beta \varphi_i] \right.\\
  &\left. \left. -2\beta^5\sech[\beta \varphi_i]^{2}\tanh[\beta \varphi_i]^3 \right)\right]
  \end{aligned}
\end{equation}
due to the property of Kronecker delta only $\Lambda$'s whose first and second indices are
coincident contributes to Eq(\ref{eq:mode2pi}). $\Lambda_{mmj}$ can be obtained by solving
Bathe-Salpeter equation:
\begin{equation}
  \Lambda_{iim}=-\sum_{k,j} G_{ik}\frac{\delta \Gamma_{kj}}{\delta \varphi_{m}}G_{ji}
\end{equation}
here
\begin{equation}
  \begin{aligned}
    \frac{\delta \Gamma_{ij}}{\delta \varphi_m}=&
\frac{\delta}{\delta \varphi_m}\left(\frac{\delta^2\Gamma[\varphi,G]}{\delta\varphi_i\delta \varphi_j}\right)_{G=\tilde{G}[\varphi]} \\
=&-\beta^{2}\frac{\delta  \text{sech}\left[\beta\varphi_{ {i}}\right]^{2}}{\delta\varphi_{m}}\delta_{ {i}{j}}
+\beta^{3}G_{ii}\frac{\delta\left(\beta  \text{sech}\left[\beta\varphi_{ {i}}\right]^{4}-2\beta
\text{sech}\left[\beta\varphi_{ {i}}\right]^{2}\tanh\left[\beta\varphi_{ {i}}\right]^{2}\right)}{\delta \varphi_m}
\delta_{ {i}{j}}\\
&+\beta^{3}\Lambda_{iim}\left(\beta  \text{sech}\left[\beta\varphi_{ {i}}\right]^{4}-2\beta
  \text{sech}\left[\beta\varphi_{ {i}}\right]^{2}\tanh\left[\beta\varphi_{ {i}}\right]^{2}\right)\delta_{ {i}{j}}
  \end{aligned}
\end{equation}

In Fourier space the external
Green's function can be written as
\begin{equation} \label{eq:mod2pi}
  \begin{aligned}
    G^{-1}_{\text{ext}}(k)=& G^{-1}(k)
    +\left[(\beta^3 \sech[\beta \varphi]^2\tanh[\beta\varphi]) \right. \\
    &\left. -G_{ii}\left(4\beta^5\sech[\beta \varphi]^{4}\tanh[\beta \varphi]
-2\beta^5\sech[\beta \varphi]^{2}\tanh[\beta \varphi]^3 \right) \right] \Lambda(k)
\\
&-\beta^6G_{ii}G_{ii}[2\text{sech}^6(\beta\varphi)
-11\text{sech}^4(\beta\varphi)\tanh^2(\beta\varphi)
+2\text{sech}^2(\beta\varphi)\tanh^4(\beta\varphi)]
\end{aligned}
\end{equation}
where $\Lambda(k)$ of 2PI method has the  same  expression
as Eq(\ref{eq:Lambda}) obtained in 1PI approach.

\section{\label{sec:res}Numerical results}
\begin{figure}[t!]
  \centering
    \includegraphics[width=0.45\textwidth,angle=0]{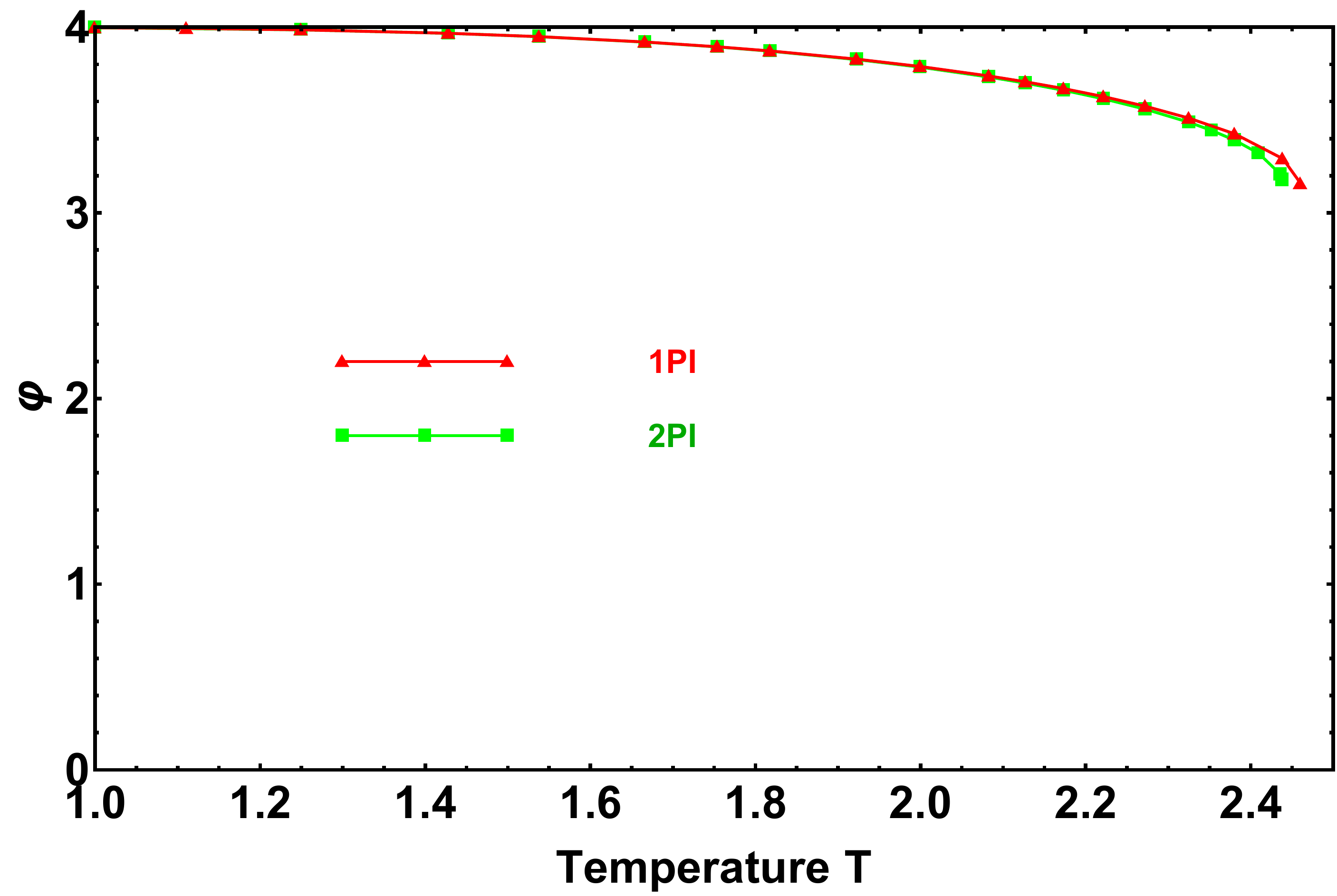}
    \caption{\label{fig:phi} $\varphi$ - $T$ for 1PI and 2PI approaches,
    $T_{c, \text{1PI}} = 2.4606$ and terminated $\varphi = 3.15737$,
    $T_{c, \text{2PI}} = 2.4390$ and $\varphi = 3.17532$}
\end{figure}

We solve 1PI equations (\ref{eq:1pishift}, \ref{eq:1pigap}), and 2PI equations
(\ref{eq:2pishift}, \ref{eq:2pigap}), respectively. The results are shown below
in $k_{B} = J = 1$ unit. $\varphi$ as a function of temperature is presented in Fig.\ref{fig:phi}.
The equation ceases to have a solution at $T_{c}$,
which is the end point of the broken phase and is actually the critical point of a
second-order phase transition. For a given $\varphi$, we can get $\langle \sigma \rangle$
from Eq(\ref{eq:defsigma}), however it is not exactly equal to the spontaneous magnetization
and needs corrections to get the exact $\langle \sigma \rangle$ just like $G$ needs corrections
to get the exact Green function.

\begin{table}
  \caption{\label{tab:tc} $T_{c}$ from various approximations and exact
  \cite{fisher1967theory, ashcroft1976solid, wysin2000correlated, zhuravlev2005molecular} values.}
  \begin{tabular}{c|c|c|c|c|c}
  \hline
  \hline
  Exact & BPW & SMF & SCCF & 1PI & 2PI \\
  \hline
  2.26918$\cdots$ & 2.885 & 2.142 & 2.595 & 2.4606 & 2.4390 \\
  \hline
  \hline
  \end{tabular}
\end{table}

In Table \ref{tab:tc} we display $T_{c}$ from 1PI, 2PI,
as well as the SCCF and SMF results \cite{wysin2000correlated, zhuravlev2005molecular},
together with either exact or approximate values from series
estimates \cite{fisher1967theory}. For the 2D square lattice
Ising model,  1PI gives $T_c = 2.4606$, and 2PI gives $T_{c} = 2.4390$, both
closer to the exact result than the BPW approximation and self-consistent
correlated field method(SCCF).

According to fluctuation-dissipation theorem, we can get the susceptibility $\chi_{ij}$ with following relation:
\begin{equation}
  \chi_{ij}=\frac{\delta \langle\sigma_i\rangle}{\delta h_j}=\beta \langle\sigma_i\sigma_j
  \rangle-\beta\langle\sigma_i\rangle\langle\sigma_j\rangle
\end{equation}

The Fourier transform of the susceptibility $\chi_{ij}$ is
\begin{equation}
  \chi_{ij}=\sum_{\alpha=x,y}\int^{\pi}_{-\pi} \frac{d^2k}{(2\pi)^2}\exp(-ik_\alpha(i_\alpha-j_\alpha))\chi(k)
\end{equation}

\begin{figure}[t!]
  \centering
      \includegraphics[width=0.45\textwidth,angle=0]{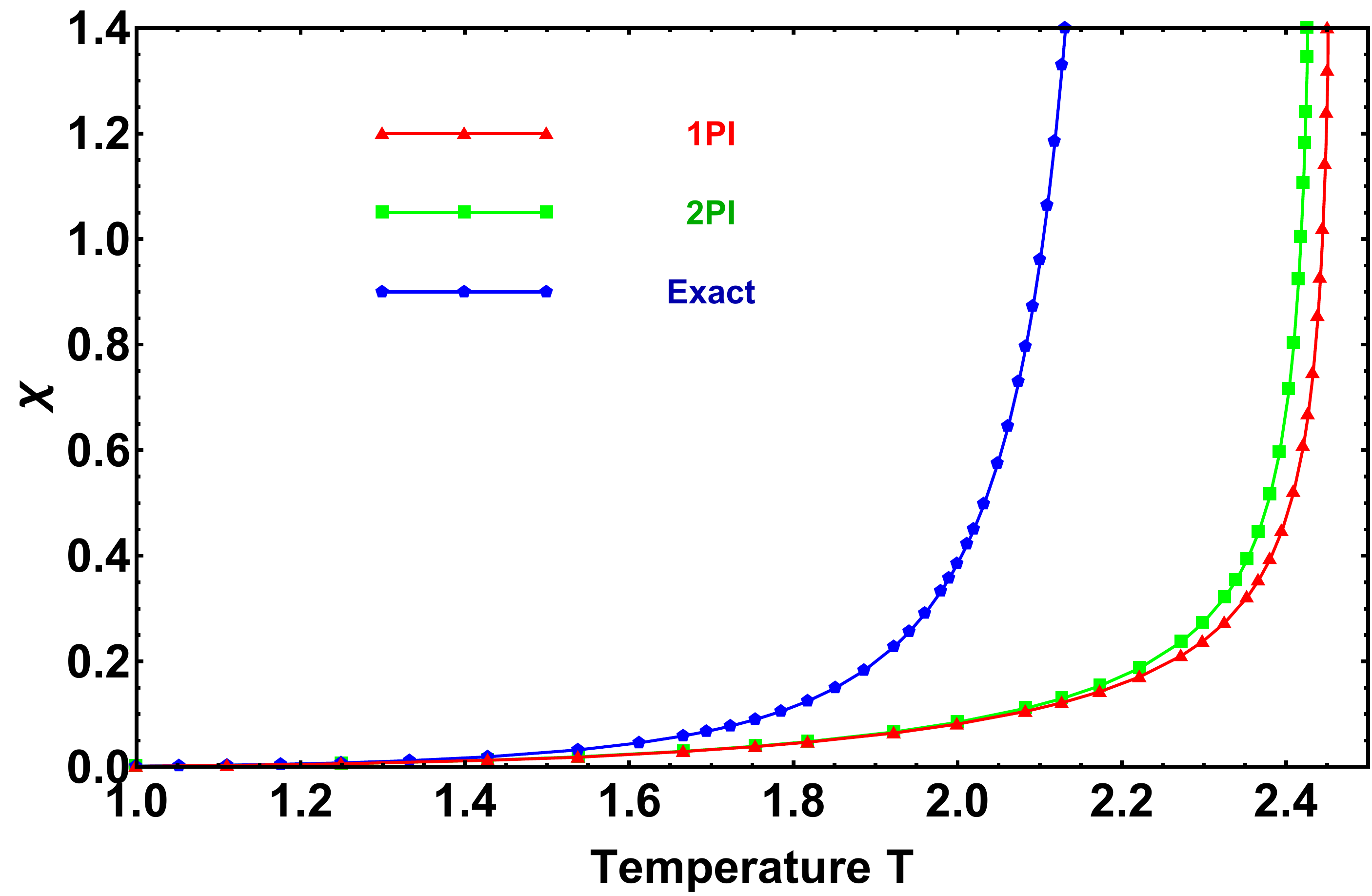}
      \caption{\label{fig:sus} $\chi$ - $T$ from 1PI, 2PI
      approaches and exact value from low temperature expansion series\cite{au2002correlation,
      orrick2001susceptibility}.}
\end{figure}

The zero momentum susceptibility, which we denote as $\chi$, can be obtained from Eq(\ref{eq:fullcorr}, \ref{eq:mod2pi})
with the following expression:

\begin{equation}
  \begin{aligned}
    \chi=&\beta\sum_m\left(\langle\sigma_m\sigma_0\rangle-\langle\sigma_0\rangle^2\right)
    =\beta\sum_m\left(J^{-1}_{m,i}J^{-1}_{0,j}G_{ij}-\beta^{-1}J^{-1}_{m,0}\right)\\
    =&\beta\left(J^{-2}(0)G(0)-\beta^{-1}J^{-1}(0)\right)
  \end{aligned}
\end{equation}
Here $G(0)$ refers to $G_{\text{full}}(0)$ or $G_{\text{ext}}(0)$, under 1PI or 2PI approximation, respectively.
The numerical results are plotted in Fig.\ref{fig:sus}, both susceptibility will diverge at its corresponding $T_{c}$.

\begin{figure}[t!]
    \includegraphics[width=0.45\textwidth,angle=0]{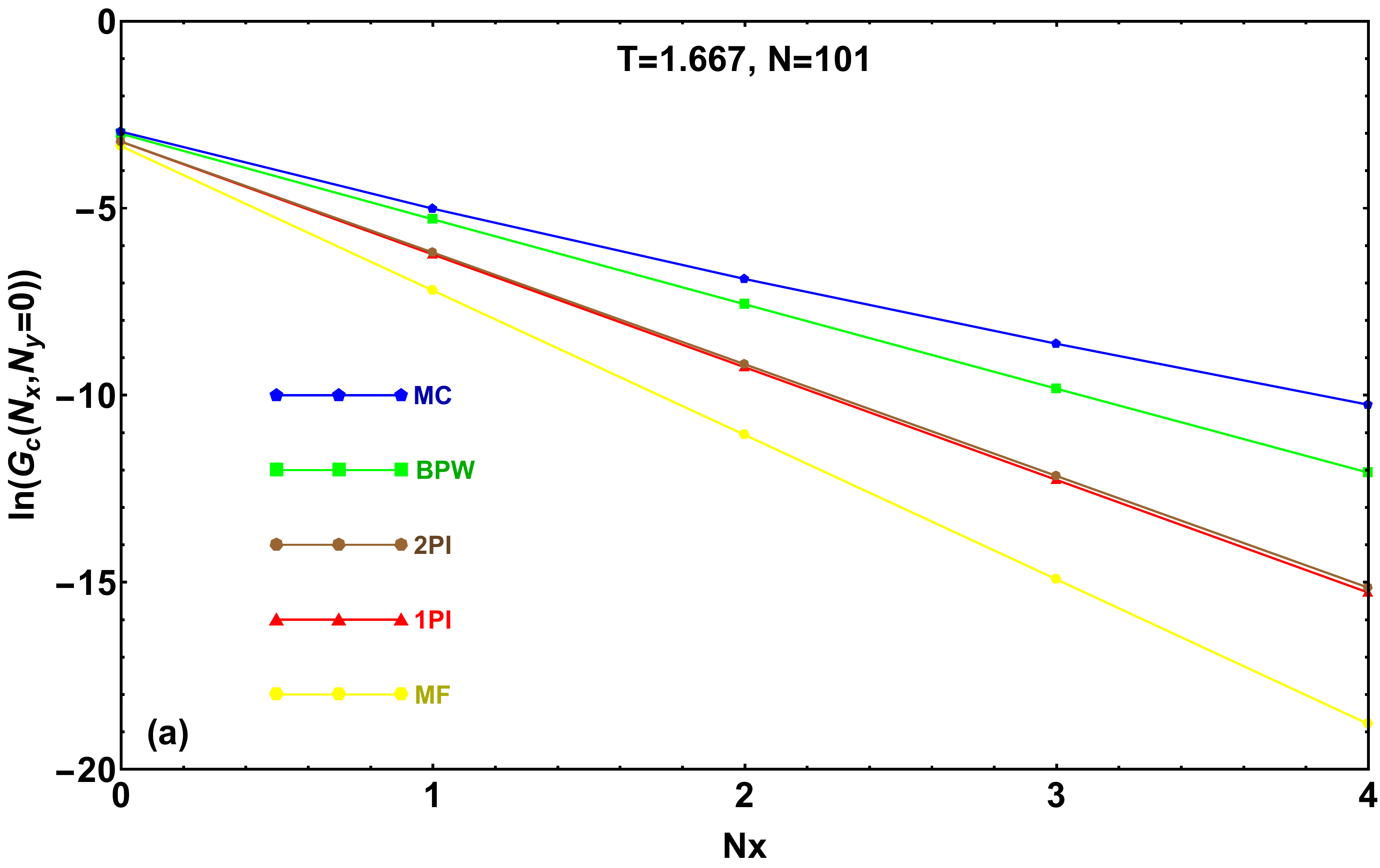}
    \includegraphics[width=0.45\textwidth,angle=0]{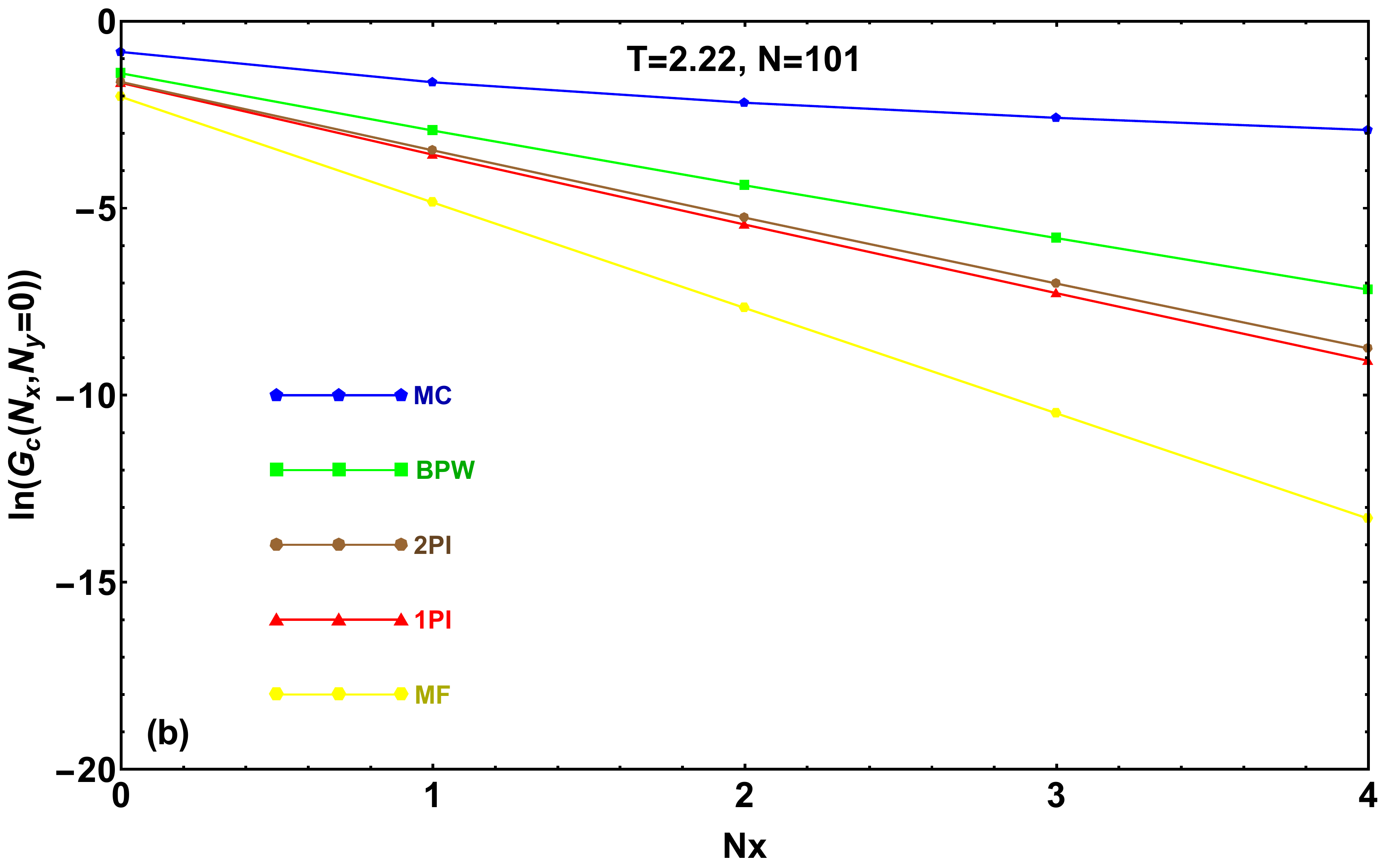}
  \caption{\label{fig:GF} The logarithm of Green's function at (a)$T = 1.667$, (b)$T = 2.22$ for $101 \times 101$ for square lattice along the x direction.
    The regular pentagons are results  by Monte Carlo simulation, green squares are BPW results (\cite{ricci2012bethe} and references therein), red triangles and brown septangles are 1PI and 2PI results respectively. The results from mean-field approach are denoted  by yellow hexagons.}
\end{figure}

We also compare results for finite size lattices (subject to periodic boundary conditions) with Monte Carlo results. And the results are illustrated in Fig.\ref{fig:GF}, under two different temperatures. For the finite size lattice of $N\times N$ with periodic boundary conditions, the
formula in the integration shall be substituted as
\begin{equation}
\frac{1}{\left( 2\pi \right) ^{2}}\int_{-\pi }^{\pi
}dk_{x}\int_{-\pi}^{\pi }dk_{y}f\left( k_{x},k_{y}\right)
\rightarrow \frac{1}{N^{2}}\sum\limits_{k_{x},k_{y}}f\left(
k_{x},k_{y}\right)
\end{equation}%
where  $f\left( k_{x},k_{y}\right) $ is a periodic function of $k_{x}$ and $%
k_{y}$ (periodicity is $2\pi $), and inside the summation, $k_{x}=%
\frac{2\pi }{N}i,i=0,N-1,k_{y}=\frac{2\pi }{N}j,j=0,N-1$  due to periodic
boundary conditions.

It can be seen that our results show a significant improvement compared to the Mean-field approach, especially for a temperature closer to $T_{c}$.
We also include the correlation function obtained by BPW method (or Bethe approximation) in Fig. 3 for comparison.  The BPW calculation of the correlation is highly non trivial and complex. It
was only studied quite recently, see \cite{ricci2012bethe} and references therein. BPW method was particular useful for studying Ising model (also useful for Random Ising model), however the generalizations to other models are too complex. BPW result is better than the field theoretical result for the correlation function below the real critical temperature ($T_{c}=2.26918$), however
the critical temperature obtained by BPW method is $T_{c}=2.885$, worse than the critical temperature obtained by the field theoretical method (for 1PI,$T_{c}=2.4606$, and for 2PI, $T_{c}=2.4390$ ).

 Although  the deviation of our approach for the correlation function with respect to MC is slightly larger than BPW method below the critical temperature, however, the field theoretical approach can easily generalize to quantum many body theory, and can be  also applied to complicated spin models with continue symmetry , for example XY model and Heisenberg model, etc.

\section{\label{sec:con}conclusion}
In conclusion, we calculate the critical temperature, susceptibility and Green
function nonperturbatively with two kind field theories developed by the Dyson-Schwinger
theory and the $\Phi$-derivable theory at leading order fluctuation correction. With relative low cost, both method are able to give fairly good predictions of $T_c$ for the Ising model.
In the area far away from critical point, the susceptibility and Green function
obtained from our method is quite accurate comparing with exact solution. This is
a systemic approach which can be used to treat more complex spin models. The methods will preserve fundamental identities, like the fluctuation dissipation relation and WTI identities for systems with continue symmetries, which are very crucial for giving consistent descriptions of such systems.

\begin{acknowledgments}
We thank Professor Baruch Rosenstein for valuable discussions.  The work is supported by National
Natural Science Foundation of China, Grants No. 11674007, No. 91736208 and No. 11920101004.
The work is also supported by High-performance Computing Platform of Peking University.
\end{acknowledgments}

\nocite{*}

\bibliography{Ising}

\begin{thebibliography}{24}%
\makeatletter
\providecommand \@ifxundefined [1]{%
 \@ifx{#1\undefined}
}%
\providecommand \@ifnum [1]{%
 \ifnum #1\expandafter \@firstoftwo
 \else \expandafter \@secondoftwo
 \fi
}%
\providecommand \@ifx [1]{%
 \ifx #1\expandafter \@firstoftwo
 \else \expandafter \@secondoftwo
 \fi
}%
\providecommand \natexlab [1]{#1}%
\providecommand \enquote  [1]{``#1''}%
\providecommand \bibnamefont  [1]{#1}%
\providecommand \bibfnamefont [1]{#1}%
\providecommand \citenamefont [1]{#1}%
\providecommand \href@noop [0]{\@secondoftwo}%
\providecommand \href [0]{\begingroup \@sanitize@url \@href}%
\providecommand \@href[1]{\@@startlink{#1}\@@href}%
\providecommand \@@href[1]{\endgroup#1\@@endlink}%
\providecommand \@sanitize@url [0]{\catcode `\\12\catcode `\$12\catcode
  `\&12\catcode `\#12\catcode `\^12\catcode `\_12\catcode `\%12\relax}%
\providecommand \@@startlink[1]{}%
\providecommand \@@endlink[0]{}%
\providecommand \url  [0]{\begingroup\@sanitize@url \@url }%
\providecommand \@url [1]{\endgroup\@href {#1}{\urlprefix }}%
\providecommand \urlprefix  [0]{URL }%
\providecommand \Eprint [0]{\href }%
\providecommand \doibase [0]{https://doi.org/}%
\providecommand \selectlanguage [0]{\@gobble}%
\providecommand \bibinfo  [0]{\@secondoftwo}%
\providecommand \bibfield  [0]{\@secondoftwo}%
\providecommand \translation [1]{[#1]}%
\providecommand \BibitemOpen [0]{}%
\providecommand \bibitemStop [0]{}%
\providecommand \bibitemNoStop [0]{.\EOS\space}%
\providecommand \EOS [0]{\spacefactor3000\relax}%
\providecommand \BibitemShut  [1]{\csname bibitem#1\endcsname}%
\let\auto@bib@innerbib\@empty
\bibitem [{\citenamefont {Ising}(1925)}]{ising1925beitrag}%
  \BibitemOpen
  \bibfield  {author} {\bibinfo {author} {\bibfnamefont {E.}~\bibnamefont
  {Ising}},\ }\bibfield  {title} {\bibinfo {title} {Beitrag zur theorie des
  ferromagnetismus},\ }\href@noop {} {\bibfield  {journal} {\bibinfo  {journal}
  {Zeitschrift f{\"u}r Physik}\ }\textbf {\bibinfo {volume} {31}},\ \bibinfo
  {pages} {253} (\bibinfo {year} {1925})}\BibitemShut {NoStop}%
\bibitem [{\citenamefont {Onsager}(1944)}]{onsager1944crystal}%
  \BibitemOpen
  \bibfield  {author} {\bibinfo {author} {\bibfnamefont {L.}~\bibnamefont
  {Onsager}},\ }\bibfield  {title} {\bibinfo {title} {Crystal statistics. i. a
  two-dimensional model with an order-disorder transition},\ }\href@noop {}
  {\bibfield  {journal} {\bibinfo  {journal} {Physical Review}\ }\textbf
  {\bibinfo {volume} {65}},\ \bibinfo {pages} {117} (\bibinfo {year}
  {1944})}\BibitemShut {NoStop}%
\bibitem [{\citenamefont {Kuzemsky}(2009)}]{kuzemsky2009statistical}%
  \BibitemOpen
  \bibfield  {author} {\bibinfo {author} {\bibfnamefont {A.}~\bibnamefont
  {Kuzemsky}},\ }\bibfield  {title} {\bibinfo {title} {Statistical mechanics
  and the physics of many-particle model systems},\ }\href@noop {} {\bibfield
  {journal} {\bibinfo  {journal} {Physics of Particles and Nuclei}\ }\textbf
  {\bibinfo {volume} {40}},\ \bibinfo {pages} {949} (\bibinfo {year}
  {2009})}\BibitemShut {NoStop}%
\bibitem [{\citenamefont {Weiss}\ \emph {et~al.}(1907)\citenamefont {Weiss},
  \citenamefont {Weiss},\ and\ \citenamefont {Stoner}}]{weiss1907magnetism}%
  \BibitemOpen
  \bibfield  {author} {\bibinfo {author} {\bibfnamefont {P.}~\bibnamefont
  {Weiss}}, \bibinfo {author} {\bibfnamefont {P.}~\bibnamefont {Weiss}},\ and\
  \bibinfo {author} {\bibfnamefont {E.}~\bibnamefont {Stoner}},\ }\bibfield
  {title} {\bibinfo {title} {Magnetism ond atomic structure},\ }\href@noop {}
  {\bibfield  {journal} {\bibinfo  {journal} {J. phys}\ }\textbf {\bibinfo
  {volume} {6}},\ \bibinfo {pages} {667} (\bibinfo {year} {1907})}\BibitemShut
  {NoStop}%
\bibitem [{\citenamefont {Wysin}\ and\ \citenamefont
  {Kaplan}(2000)}]{wysin2000correlated}%
  \BibitemOpen
  \bibfield  {author} {\bibinfo {author} {\bibfnamefont {G.}~\bibnamefont
  {Wysin}}\ and\ \bibinfo {author} {\bibfnamefont {J.}~\bibnamefont {Kaplan}},\
  }\bibfield  {title} {\bibinfo {title} {Correlated molecular-field theory for
  ising models},\ }\href@noop {} {\bibfield  {journal} {\bibinfo  {journal}
  {Physical Review E}\ }\textbf {\bibinfo {volume} {61}},\ \bibinfo {pages}
  {6399} (\bibinfo {year} {2000})}\BibitemShut {NoStop}%
\bibitem [{\citenamefont {Bethe}(1935)}]{bethe1935statistical}%
  \BibitemOpen
  \bibfield  {author} {\bibinfo {author} {\bibfnamefont {H.~A.}\ \bibnamefont
  {Bethe}},\ }\bibfield  {title} {\bibinfo {title} {Statistical theory of
  superlattices},\ }\href@noop {} {\bibfield  {journal} {\bibinfo  {journal}
  {Proceedings of the Royal Society of London. Series A-Mathematical and
  Physical Sciences}\ }\textbf {\bibinfo {volume} {150}},\ \bibinfo {pages}
  {552} (\bibinfo {year} {1935})}\BibitemShut {NoStop}%
\bibitem [{\citenamefont {Peierls}(1936)}]{peierls1936ising}%
  \BibitemOpen
  \bibfield  {author} {\bibinfo {author} {\bibfnamefont {R.}~\bibnamefont
  {Peierls}},\ }\bibfield  {title} {\bibinfo {title} {On ising's model of
  ferromagnetism},\ }in\ \href@noop {} {\emph {\bibinfo {booktitle}
  {Mathematical Proceedings of the Cambridge Philosophical Society}}},\
  Vol.~\bibinfo {volume} {32}\ (\bibinfo {organization} {Cambridge University
  Press},\ \bibinfo {year} {1936})\ pp.\ \bibinfo {pages}
  {477--481}\BibitemShut {NoStop}%
\bibitem [{\citenamefont {Weiss}(1948)}]{weiss1948application}%
  \BibitemOpen
  \bibfield  {author} {\bibinfo {author} {\bibfnamefont {P.~R.}\ \bibnamefont
  {Weiss}},\ }\bibfield  {title} {\bibinfo {title} {The application of the
  bethe-peierls method to ferromagnetism},\ }\href@noop {} {\bibfield
  {journal} {\bibinfo  {journal} {Physical Review}\ }\textbf {\bibinfo {volume}
  {74}},\ \bibinfo {pages} {1493} (\bibinfo {year} {1948})}\BibitemShut
  {NoStop}%
\bibitem [{\citenamefont {Zhuravlev}(2005)}]{zhuravlev2005molecular}%
  \BibitemOpen
  \bibfield  {author} {\bibinfo {author} {\bibfnamefont {K.~K.}\ \bibnamefont
  {Zhuravlev}},\ }\bibfield  {title} {\bibinfo {title} {Molecular-field theory
  method for evaluating critical points of the ising model},\ }\href@noop {}
  {\bibfield  {journal} {\bibinfo  {journal} {Physical Review E}\ }\textbf
  {\bibinfo {volume} {72}},\ \bibinfo {pages} {056104} (\bibinfo {year}
  {2005})}\BibitemShut {NoStop}%
\bibitem [{\citenamefont {Yamamoto}(2009)}]{yamamoto2009correlated}%
  \BibitemOpen
  \bibfield  {author} {\bibinfo {author} {\bibfnamefont {D.}~\bibnamefont
  {Yamamoto}},\ }\bibfield  {title} {\bibinfo {title} {Correlated cluster
  mean-field theory for spin systems},\ }\href@noop {} {\bibfield  {journal}
  {\bibinfo  {journal} {Physical Review B}\ }\textbf {\bibinfo {volume} {79}},\
  \bibinfo {pages} {144427} (\bibinfo {year} {2009})}\BibitemShut {NoStop}%
\bibitem [{\citenamefont {Viana}\ \emph {et~al.}(2014)\citenamefont {Viana},
  \citenamefont {Salmon}, \citenamefont {de~Sousa}, \citenamefont {Neto},\ and\
  \citenamefont {Padilha}}]{viana2014effective}%
  \BibitemOpen
  \bibfield  {author} {\bibinfo {author} {\bibfnamefont {J.~R.}\ \bibnamefont
  {Viana}}, \bibinfo {author} {\bibfnamefont {O.~R.}\ \bibnamefont {Salmon}},
  \bibinfo {author} {\bibfnamefont {J.~R.}\ \bibnamefont {de~Sousa}}, \bibinfo
  {author} {\bibfnamefont {M.~A.}\ \bibnamefont {Neto}},\ and\ \bibinfo
  {author} {\bibfnamefont {I.~T.}\ \bibnamefont {Padilha}},\ }\bibfield
  {title} {\bibinfo {title} {An effective correlated mean-field theory applied
  in the spin-1/2 ising ferromagnetic model},\ }\href@noop {} {\bibfield
  {journal} {\bibinfo  {journal} {Journal of magnetism and magnetic materials}\
  }\textbf {\bibinfo {volume} {369}},\ \bibinfo {pages} {101} (\bibinfo {year}
  {2014})}\BibitemShut {NoStop}%
\bibitem [{\citenamefont {Luttinger}\ and\ \citenamefont
  {Ward}(1960)}]{luttinger1960ground}%
  \BibitemOpen
  \bibfield  {author} {\bibinfo {author} {\bibfnamefont {J.~M.}\ \bibnamefont
  {Luttinger}}\ and\ \bibinfo {author} {\bibfnamefont {J.~C.}\ \bibnamefont
  {Ward}},\ }\bibfield  {title} {\bibinfo {title} {Ground-state energy of a
  many-fermion system. ii},\ }\href@noop {} {\bibfield  {journal} {\bibinfo
  {journal} {Physical Review}\ }\textbf {\bibinfo {volume} {118}},\ \bibinfo
  {pages} {1417} (\bibinfo {year} {1960})}\BibitemShut {NoStop}%
\bibitem [{\citenamefont {Baym}\ and\ \citenamefont
  {Kadanoff}(1961)}]{baym1961conservation}%
  \BibitemOpen
  \bibfield  {author} {\bibinfo {author} {\bibfnamefont {G.}~\bibnamefont
  {Baym}}\ and\ \bibinfo {author} {\bibfnamefont {L.~P.}\ \bibnamefont
  {Kadanoff}},\ }\bibfield  {title} {\bibinfo {title} {Conservation laws and
  correlation functions},\ }\href@noop {} {\bibfield  {journal} {\bibinfo
  {journal} {Physical Review}\ }\textbf {\bibinfo {volume} {124}},\ \bibinfo
  {pages} {287} (\bibinfo {year} {1961})}\BibitemShut {NoStop}%
\bibitem [{\citenamefont {Cornwall}\ \emph {et~al.}(1974)\citenamefont
  {Cornwall}, \citenamefont {Jackiw},\ and\ \citenamefont
  {Tomboulis}}]{cornwall1974effective}%
  \BibitemOpen
  \bibfield  {author} {\bibinfo {author} {\bibfnamefont {J.~M.}\ \bibnamefont
  {Cornwall}}, \bibinfo {author} {\bibfnamefont {R.}~\bibnamefont {Jackiw}},\
  and\ \bibinfo {author} {\bibfnamefont {E.}~\bibnamefont {Tomboulis}},\
  }\bibfield  {title} {\bibinfo {title} {Effective action for composite
  operators},\ }\href@noop {} {\bibfield  {journal} {\bibinfo  {journal}
  {Physical Review D}\ }\textbf {\bibinfo {volume} {10}},\ \bibinfo {pages}
  {2428} (\bibinfo {year} {1974})}\BibitemShut {NoStop}%
\bibitem [{\citenamefont {Kovner}\ and\ \citenamefont
  {Rosenstein}(1989)}]{kovner1989covariant}%
  \BibitemOpen
  \bibfield  {author} {\bibinfo {author} {\bibfnamefont {A.}~\bibnamefont
  {Kovner}}\ and\ \bibinfo {author} {\bibfnamefont {B.}~\bibnamefont
  {Rosenstein}},\ }\bibfield  {title} {\bibinfo {title} {Covariant gaussian
  approximation. i. formalism},\ }\href@noop {} {\bibfield  {journal} {\bibinfo
   {journal} {Physical Review D}\ }\textbf {\bibinfo {volume} {39}},\ \bibinfo
  {pages} {2332} (\bibinfo {year} {1989})}\BibitemShut {NoStop}%
\bibitem [{\citenamefont {Van~Hees}\ and\ \citenamefont
  {Knoll}(2002)}]{van2002renormalization}%
  \BibitemOpen
  \bibfield  {author} {\bibinfo {author} {\bibfnamefont {H.}~\bibnamefont
  {Van~Hees}}\ and\ \bibinfo {author} {\bibfnamefont {J.}~\bibnamefont
  {Knoll}},\ }\bibfield  {title} {\bibinfo {title} {Renormalization in
  self-consistent approximation schemes at finite temperature. iii. global
  symmetries},\ }\href@noop {} {\bibfield  {journal} {\bibinfo  {journal}
  {Physical Review D}\ }\textbf {\bibinfo {volume} {66}},\ \bibinfo {pages}
  {025028} (\bibinfo {year} {2002})}\BibitemShut {NoStop}%
\bibitem [{\citenamefont {Amit}\ and\ \citenamefont
  {Martin-Mayor}(2005)}]{amit2005field}%
  \BibitemOpen
  \bibfield  {author} {\bibinfo {author} {\bibfnamefont {D.~J.}\ \bibnamefont
  {Amit}}\ and\ \bibinfo {author} {\bibfnamefont {V.}~\bibnamefont
  {Martin-Mayor}},\ }\href@noop {} {\emph {\bibinfo {title} {Field theory, the
  renormalization group, and critical phenomena: graphs to computers}}}\
  (\bibinfo  {publisher} {World Scientific Publishing Company},\ \bibinfo
  {year} {2005})\BibitemShut {NoStop}%
\bibitem [{\citenamefont {Wang}\ \emph {et~al.}(2017)\citenamefont {Wang},
  \citenamefont {Li}, \citenamefont {Kao},\ and\ \citenamefont
  {Rosenstein}}]{wang2017covariant}%
  \BibitemOpen
  \bibfield  {author} {\bibinfo {author} {\bibfnamefont {J.}~\bibnamefont
  {Wang}}, \bibinfo {author} {\bibfnamefont {D.}~\bibnamefont {Li}}, \bibinfo
  {author} {\bibfnamefont {H.}~\bibnamefont {Kao}},\ and\ \bibinfo {author}
  {\bibfnamefont {B.}~\bibnamefont {Rosenstein}},\ }\bibfield  {title}
  {\bibinfo {title} {Covariant gaussian approximation in ginzburg--landau
  model},\ }\href@noop {} {\bibfield  {journal} {\bibinfo  {journal} {Annals of
  Physics}\ }\textbf {\bibinfo {volume} {380}},\ \bibinfo {pages} {228}
  (\bibinfo {year} {2017})}\BibitemShut {NoStop}%
\bibitem [{\citenamefont {Rosenstein}\ and\ \citenamefont
  {Kovner}(1989)}]{rosenstein1989covariant}%
  \BibitemOpen
  \bibfield  {author} {\bibinfo {author} {\bibfnamefont {B.}~\bibnamefont
  {Rosenstein}}\ and\ \bibinfo {author} {\bibfnamefont {A.}~\bibnamefont
  {Kovner}},\ }\bibfield  {title} {\bibinfo {title} {Covariant gaussian
  approximation. ii. scalar theories},\ }\href@noop {} {\bibfield  {journal}
  {\bibinfo  {journal} {Physical Review D}\ }\textbf {\bibinfo {volume} {40}},\
  \bibinfo {pages} {504} (\bibinfo {year} {1989})}\BibitemShut {NoStop}%
\bibitem [{\citenamefont {Fisher}(1967)}]{fisher1967theory}%
  \BibitemOpen
  \bibfield  {author} {\bibinfo {author} {\bibfnamefont {M.~E.}\ \bibnamefont
  {Fisher}},\ }\bibfield  {title} {\bibinfo {title} {The theory of equilibrium
  critical phenomena},\ }\href@noop {} {\bibfield  {journal} {\bibinfo
  {journal} {Reports on progress in physics}\ }\textbf {\bibinfo {volume}
  {30}},\ \bibinfo {pages} {615} (\bibinfo {year} {1967})}\BibitemShut
  {NoStop}%
\bibitem [{\citenamefont {Ashcroft}\ \emph {et~al.}(1976)\citenamefont
  {Ashcroft}, \citenamefont {Mermin} \emph {et~al.}}]{ashcroft1976solid}%
  \BibitemOpen
  \bibfield  {author} {\bibinfo {author} {\bibfnamefont {N.~W.}\ \bibnamefont
  {Ashcroft}}, \bibinfo {author} {\bibfnamefont {N.~D.}\ \bibnamefont
  {Mermin}}, \emph {et~al.},\ }\href@noop {} {\emph {\bibinfo {title} {Solid
  state physics}}},\ Vol.\ \bibinfo {volume} {2005}\ (\bibinfo  {publisher}
  {holt, rinehart and winston, new york London},\ \bibinfo {year}
  {1976})\BibitemShut {NoStop}%
\bibitem [{\citenamefont {Au-Yang}\ and\ \citenamefont
  {Perk}(2002)}]{au2002correlation}%
  \BibitemOpen
  \bibfield  {author} {\bibinfo {author} {\bibfnamefont {H.}~\bibnamefont
  {Au-Yang}}\ and\ \bibinfo {author} {\bibfnamefont {J.~H.}\ \bibnamefont
  {Perk}},\ }\bibfield  {title} {\bibinfo {title} {Correlation functions and
  susceptibility in the z-invariant ising model},\ }in\ \href@noop {} {\emph
  {\bibinfo {booktitle} {MathPhys Odyssey 2001}}}\ (\bibinfo  {publisher}
  {Springer},\ \bibinfo {year} {2002})\ pp.\ \bibinfo {pages}
  {23--48}\BibitemShut {NoStop}%
\bibitem [{\citenamefont {Orrick}\ \emph {et~al.}(2001)\citenamefont {Orrick},
  \citenamefont {Nickel}, \citenamefont {Guttmann},\ and\ \citenamefont
  {Perk}}]{orrick2001susceptibility}%
  \BibitemOpen
  \bibfield  {author} {\bibinfo {author} {\bibfnamefont {W.}~\bibnamefont
  {Orrick}}, \bibinfo {author} {\bibfnamefont {B.}~\bibnamefont {Nickel}},
  \bibinfo {author} {\bibfnamefont {A.}~\bibnamefont {Guttmann}},\ and\
  \bibinfo {author} {\bibfnamefont {J.}~\bibnamefont {Perk}},\ }\bibfield
  {title} {\bibinfo {title} {The susceptibility of the square lattice ising
  model: new developments},\ }\href@noop {} {\bibfield  {journal} {\bibinfo
  {journal} {Journal of Statistical Physics}\ }\textbf {\bibinfo {volume}
  {102}},\ \bibinfo {pages} {795} (\bibinfo {year} {2001})},\ \bibinfo {note}
  {for the complete set of series coefficients see
  https://blogs.unimelb.edu.au/tony-guttmann/.}\BibitemShut {Stop}%
\bibitem [{\citenamefont {Ricci-Tersenghi}(2012)}]{ricci2012bethe}%
  \BibitemOpen
  \bibfield  {author} {\bibinfo {author} {\bibfnamefont {F.}~\bibnamefont
  {Ricci-Tersenghi}},\ }\bibfield  {title} {\bibinfo {title} {The bethe
  approximation for solving the inverse ising problem: a comparison with other
  inference methods},\ }\href@noop {} {\bibfield  {journal} {\bibinfo
  {journal} {Journal of Statistical Mechanics: Theory and Experiment}\ }\textbf
  {\bibinfo {volume} {2012}},\ \bibinfo {pages} {P08015} (\bibinfo {year}
  {2012})}\BibitemShut {NoStop}%
\end{thebibliography}%

\end{document}